\documentclass[12pt,preprint]{aastex}  

\def\swift{{\it Swift}}

\def\sax{{\it BeppoSAX}}

\def\fermi{{\it Fermi}}  
  
\def\lum{erg s$^{-1}$}  
  
\def\nh{cm$^{-2}$}  
  
\def\deg{$^{\circ}$}  
  
\def\ltsima{$\; \buildrel < \over \sim \;$}  
\def\simlt{\lower.5ex\hbox{\ltsima}} 
\def\gtsima{$\; \buildrel > \over \sim \;$}  
\def\simgt{\lower.5ex\hbox{\gtsima}} 
  
\def\3c{3C~390.3}

\input{psfig.sty}

\begin{document}  
  
\title{\swift\ BAT, \fermi\ LAT, and the Blazar Sequence}

\author{R. M. Sambruna}
\affil{NASA/GSFC, Code 662, Greenbelt, MD 20771}  
  
\author{D. Donato}
\affil{NASA/GSFC, Code 661, Greenbelt, MD 20771}  

\author{M. Ajello} \affil{SLAC National Laboratory and Kavli Institute
  for Particle Astrophysics and Cosmology, 2575 Sand Hill Road, Menlo
  Park, CA 94025, USA}
  
\author{L. Maraschi} \affil{INAF - Osservatorio Astronomico di Brera,
  V. Brera 28, I-20100 Milano, Italy}
  
\author{J. Tueller, W. Baumgartner, G. Skinner, C. Markwardt, S. Barthelmy, 
 N. Gehrels} 
\affil{NASA/GSFC, Code 661, Greenbelt, MD 20771}
  
\author{R.F.Mushotzky}
\affil{NASA/GSFC, Code 662, Greenbelt, MD 20771}

\begin{abstract}

Using public \fermi\ LAT and \swift\ BAT observations, we constructed
the first sample of blazars selected at both hard X-rays and
gamma-rays. Studying its spectral properties, we find a luminosity
dependence of the spectral slopes at both energies.  Specifically,
luminous blazars, generally classified as FSRQs, have {\it hard}
continua in the medium-hard X-ray range but {\it soft} continua in the
LAT gamma-ray range (photon indices $\Gamma_X$ \ltsima 2 and
$\Gamma_G$ \gtsima 2), while lower luminosity blazars, classified as
BL~Lacs, have opposite behavior, i.e., {\it soft} X-ray and {\it hard}
gamma-ray continua ($\Gamma_X$ \gtsima 2.4 and $\Gamma_G < 2$). The
trends are confirmed by detailed Monte Carlo simulations explicitly
taking into account the observational biases of both instruments. Our
results support the so-called ``blazar sequence'' which was originally
based on radio samples of blazars and radio luminosities. We also
argue that the X-ray-to-gamma-ray continua of blazars may provide
independent insights into the physical conditions around the jet,
complementing/superseding the ambiguities of the traditional
classification based on optical properties.

\end{abstract}

{\sl Subject Headings:} {Galaxies: active --- galaxies: radio --
  galaxies: individual --- X-rays: galaxies}

\section{Introduction}  
  
Our study of the high-energy extragalactic sky has been propelled
forward thanks to the launch in June 2008 of the \fermi\ Gamma-ray
Observatory. After only three months of operations the LAT experiment
already detected over a hundred sources at \gtsima $10\sigma$, of
which 99 are blazars and 2 are radio galaxies (Abdo et al. 2009a). The
publication of the LAT Bright Source Catalog (Abdo et al. 2009a)
prompted a few studies of the statistical properties of gamma-ray
selected blazars (Ghisellini, Maraschi, \& Tavecchio, 2009; Kovalev et
al. 2009; Lister et al. 2009), as well as of their Spectral Energy
Distributions (SEDs; see e.g., Abdo et al. 2009c).

One of the hot debates concerning blazar SEDs is the general validity
of the so-called ``blazar sequence''. According to the latter, when
blazar SEDs are arranged in order of decreasing bolometric luminosity
their shapes change in an orderly fashion, with the synchrotron and
Compton peaks moving to higher energies in accord (Fossati et
al. 1998). It is important to stress that the original blazar sequence
was discovered using blazar samples selected from the radio and
X-rays, but binned according to radio luminosity only (Fossati et
al. 1998), leading to biases (Padovani 2007; but see Ghisellini \&
Tavecchio 2008, Maraschi et al. 2008 for a rebuttal). Moreover the
gamma-ray data available at the time were scarce (Maraschi \&
Tavecchio 2001). A replica of the sequence at gamma-rays was recently
found by Ghisellini, Maraschi, \& Tavecchio (2009) based on the
\fermi\ 3 months blazar list. They found that more luminous gamma-ray
sources have softer LAT spectra, with the BL Lac and FSRQ populations
occupying neatly separated regions. The dividing luminosity was
interpreted in terms of a critical accretion rate, $\sim 10^{-2}$ the
Eddington one, regulating the transition from efficient accretion in
FSRQs to inefficient accretion in BL Lacs (Ghisellini, Maraschi, \&
Tavecchio 2009).
  
With its large field of view, nearly continuous sky coverage, and
improved sensitivity (35~mCrab in 2~ks and $\sim1$~mCrab in 36 months
in 14.5--195~keV) the BAT experiment onboard \swift\ is an important
complement to \fermi\ for blazar science. Its sensitivity to flaring
hard X-ray sources, and to quiescent emission of blazars out to $z =
4$, makes it ideal for studying the high-energy SEDs in tandem with
the LAT. The BAT has already performed a 3 year survey of the sky,
detecting 38 blazars at higher galactic latitudes (Ajello et
al. 2009); using these data we constructed the first high-energy
selected sample of blazars detected at both hard X-ray and GeV
gamma-rays. Here we study the high-energy SEDs of this sample
combining the BAT and LAT spectra, showing that the main result of the
spectral sequence - a luminosity dependence of the high-energy peak -
is independently recovered.
  
This paper is structured as follows. \S~2 describes the sample.  \S~3
presents the results of the BAT-LAT comparison, and \S~4 their
interpretation. A concordance cosmology with H$_0=71$ km s$^{-1}$
Mpc$^{-1}$, $\Omega_{\Lambda}$=0.73, and $\Omega_m$=0.27 (Spergel et
al. 2003) is adopted. Photon indices are defined as $\Gamma=\alpha-1$,
where the energy index $\alpha$ is such that F$_{\nu} \propto
\nu^{-\alpha}$. Luminosities were K-corrected according to the equation 
L$_K$=L/(1+z)$^{(1-\alpha)}$.

\section{Sample and Data}  
  
The sample was derived primarily from the list of 99 blazars at $|b|
>$ 10\deg\ detected with the LAT at more than 10$\sigma$ in the first 3
months of operations during August 4 -- October 8, 2008 (Abdo et
al. 2009a).  To increase the sample size for intersection with the BAT
sample, we also considered the list of detected extragalactic sources
which include those on the Galactic plane as well (Abdo et
al. 2009b). Absorption at X-rays is not a concern, as the BAT operates
at energies \gtsima 15~keV and is thus unaffected by typical column
densities N$_H<10^{24}$ \nh.
  
BAT performed a 3 year survey of the sky detecting 38 blazars above a
significance of 5\,$\sigma$ at $|b| >$ 10\deg. As noted in Ajello et
al. (2009), only 12 of the blazars detected by BAT above 5$\sigma$
were also detected by LAT. We thus searched the BAT all-sky image for
statistical fluctuations above the 3$\sigma$ level at the positions of
the 99 LAT blazars. This produced a list of 22 LAT blazars which are
also detected by BAT above the 3\,$\sigma$ level.  The probability
that at least one of these $\geq3$\,$\sigma$ fluctuations is given by
chance is only 0.1. There is no intrinsic difference between the
samples of BAT sources selected above 3\,$\sigma$ and 5\,$\sigma$
which are common to the LAT dataset.  Indeed, the mean redshift of the
12 and 22 BAT sources (selected respectively above 5\,$\sigma$ and
3\,$\sigma$) is 0.75 and 0.73 being the dispersion of both
distributions 0.77.  The average flux of the sample decreases from
~6.7$\times 10^{-12}$\,erg cm$^{-2}$ s$^{-1}$ to 5.1$\times
10^{-12}$\,erg cm$^{-2}$ s$^{-1}$.

Thus, the final sample used in this work includes a total of 22
blazars, of which 10 are BL Lacs and 12 are FSRQs. The confirmed TeV
sources among the BL Lacs are four - Mrk~421, Mrk~501, 1ES~1959+650,
and BL~Lac itself. We stress that this is a {\it hard X-ray/gamma-ray
  selected} sample, and therefore {\it unbiased toward the radio and
  optical properties}.
  
The data used for this work are derived from the spectral
analysis. For the BAT, spectra were extracted according to the
procedure detailed elsewhere (Ajello et al. 2009, Tueller et al. 2008;
consistent with each other) and fitted with single power law models,
deriving photon indices $\Gamma_X$ and broad-band fluxes integrated in
the 14--195~keV energy range. A single power law provides a good fit
to the BAT spectra of all sources. For the LAT, Table~3 in Abdo et
al. (2009a) provides the LAT photon index $\Gamma_G$ and fluxes
derived in various ways; we choose to use the flux derived by a
power-law fit in the energy range 0.2--100~GeV. Results do not change 
if fluxes derived with other methods are used instead. 

For 1ES~0033+59.5 and PKS~1830--21, only multiband fluxes are provided
in Abdo et al. (2009b). We used these fluxes to derive a crude
estimate of the spectral index, which in both cases turned out to be
$>2$. As a sanity check, we repeated this exercise for all the sources
in common between Abdo et al. (2009a) and (2009b), and compared the
photon index derived from the flux ratios to the photon index
$\Gamma_G$, finding good agreement within 10\% especially for
$\Gamma_G$ \ltsima 2. This gives us confidence that the slopes from
the flux ratios are a sensible approximation of the LAT continua for
the two sources. 

Finally, the BAT survey overlaps with the 3-months of the LAT
operations, and thus contains simultaneous data. While in principle it
should be possible to consider only the BAT data segments overlapping
in time with the LAT data, in practice this is not feasible due to the
limited sensitivity of the detector. The fluxes and spectra analyzed
here thus represent average states integrated over the whole 36 months
of the BAT survey. For a more detailed discussion of non-simultaneous
data and the blazar sequence, see Ghisellini et al. (2009) and
references therein.

\section{Results}  
  
First, we examine the redshift distribution of the sources to ensure
that it is randomly drawn and thus representative of the whole
population.  Figure~1 shows the distribution in redshift of the
sources (shaded histogram) compared to the total LAT blazar population
from the 3-months survey.  There is no difference between the two
histograms, as confirmed by a Kolmogorov-Smirnov (KS) test, indicating
no bias in distance/volume distribution for the GeV sources. Both LAT
and BAT are sensitive to blazar emission up to large redshifts,
because of intrinsic beaming of the jet radiation (Ajello et
al. 2009). In addition, while the selection of the sample based on the
large TS value in the LAT data favors bright/flaring sources, Figure~1
stresses that there is no bias towards to most {\it luminous}
blazars. 

Figure~2 shows the plot of the LAT and BAT spectral indices. FSRQs are
plotted with open circles, while BL Lacs with open triangles. The BL
Lacs detected at TeV energies are plotted with starred triangles.
1ES~0033+59.5 and PKS~1830--21 are plotted in dotted lines (see
\S~2). In Figure~2 blazars segregate in two distinct regions of the
diagram, one with $\Gamma_X$ \ltsima 2 and $\Gamma_G$ \gtsima 2, and
the other with $\Gamma_X$ \gtsima 2.4 and $\Gamma_G < 2$. The former
region, with {\it soft} gamma-ray but {\it hard} X-ray continua, is
populated by FSRQs and some BL~Lacs; the latter, with {\it hard}
gamma-ray but {\it soft} X-ray continua, contains only BL~Lacs,
including 3 confirmed TeV sources. These soft X-ray continuum BL~Lacs
are also known as HBLs - High Energy-peaked BL Lacs. A KS test
confirms that FSRQs and HBLs are distributed differently both in
$\Gamma_G$ and $\Gamma_X$ at 3$\sigma$ or more. 

The BL Lac 1ES~0033+59.5, marked in the Figure, is an outlier being
positioned in the region of steep BAT spectra despite its LAT
continuum similar to other BL Lacs. Large X-ray variability affects
its location. To showcase this, we plotted its position in the diagram
as implied by a previous \sax\ broad-band measurement that gave a
flatter X-ray continuum than the BAT (Costamante et al. 2001; Donato
et al. 2004). 

Interestingly, within the $\Gamma_X$ \ltsima 2 and $\Gamma_G$ \gtsima
2 region there is a hint for a further separation in $\Gamma_G$
between FSRQs and BL Lacs. The KS test probability that the two
distributions are drawn from the same parent one is 6.7 $\times
10^{-4}$. These ``intermediate'' BL Lacs are also distributed
differently than HBLs, in both spectral indices.
  
Since FSRQs are distant and luminous at GeV gamma-rays while BL Lacs
tend to be closer and fainter (e.g., Ghisellini et al. 2009), Figure~2
suggests a relationship between slopes and luminosities as well.  A
trend between $\Gamma_{G}$ and the GeV luminosity is already known
(Ghisellini et al. 2009). Here we add the BAT data to show how the
shape of the high-energy component, from X-ray to gamma-rays, changes
with gamma-ray luminosity.

Figure~3a (top panel) shows the plots of the LAT photon index,
$\Gamma_G$, versus the gamma-ray luminosity while in Figure~3b (bottom 
panel) we show the plot of the BAT spectral index, $\Gamma_X$, versus
the BAT luminosity. Figure~3a is the well-known correlation from Abdo
et al. (2009a) and Ghisellini et al. (2009), showing how the LAT slope
steepens going from lower-luminosity BL~Lacs to higher-luminosity
FSRQs. The new piece of information added here is the trend in the
bottom panel, that shows an opposite trend for $\Gamma_X$ to harden
with increasing luminosity. 
  
There are two outlier BL~Lacs in Figure~3, with luminosities
$>10^{46}$ \lum\ and continuum slopes similar to FSRQs: PKS~0537--441
and PKS~0426--380. The former, at z=0.892, is variously classified as
a FSRQ/BL Lac due to its variable broad optical lines (Maraschi et
al. 1985). Thus this source can be considered a transition object
where a broad line region is present but the emission lines are
swamped at times by the variable non-thermal continuum. Similarly, a
broad H$\alpha$ line was detected on one instance in BL~Lac itself,
the prototype of the class (Vermeulen et al. 1995), which in Figure~2
is located in the region of the intermediate group (starred triangle).
The second discrepant object is PKS~0426--380 at z=1.11. In fact its
redshift was derived from a broad MgII line in emission, while
absorption redshifts due to intervening galaxies had been measured
previously (Heidt et al.  2004). Thus also for PKS~0426--380, a
radio-selected BL Lac from the Stickel et al. (1991) sample, the BL
Lac classification is questionable.
   
The results of Figure~3 are recast in different visual format in
Figure~4. Here we show the high-energy SEDs of the sources grouped in
five bins of increasing X-ray luminosity, following the procedure of
Fossati et al. (1998) at radio. Plotted in Figure~4 are the average
BAT and LAT SEDs and their 1$\sigma$ deviations. The Figure shows the
shift of the SED shape with the luminosity.

To assess the possibility of biases introduced by the limited
sensitivity of the instruments, we performed detailed Monte Carlo
simulations, assuming random distributions of gamma-ray luminosities,
and gamma-ray and X-ray spectral indices (see Appendix). The
probability of obtaining by chance the plots in Figure~3a-b is 3.4\%,
i.e., the probability that the observed trends between the luminosity
and the slopes at high energies are intrinsic is 96.6\%. We thus
conclude that there is a trend in Figure~3 and 4 for the shape of the
high-energy SED to change with gamma-ray luminosity, specifically, its
peak shifts forward with decreasing luminosity.

\section{Discussion and Conclusions}  

We have examined the hard X-ray and gamma-ray spectral properties of a
sample of blazars selected at high energies. Our main observational
findings, summarized by Figures~2 and 3, are as follows: 1) FSRQs and
HBLs segregate apart in the shape of their high-energy SEDs.  FSRQs
have soft slopes in the gamma-ray band (100~MeV--300~GeV) and hard
slopes in the X-ray band (15--195~keV), and HBLs have hard continua
at gamma-rays and soft continua at X-rays. The remaining BL~Lacs form
a transition group between the two; and 2) this division in slopes
depends on gamma-ray luminosity. In other words, the hard X-ray and
gamma-ray spectra jointly can univocally determine the location of the
high-energy peak in the SED.
  
We interpret these results in the context of the double-humped blazar
SEDs mentioned in \S~1. First, it is important to visualize where the
sensitivity energy ranges of the BAT and LAT are located with respect
to the SEDs. In FSRQs, the BAT energy range covers the region where
the inverse Compton emission starts to emerge, while the LAT spans the
Compton peak and above; thus, we expect to observe hard X-ray spectra
and soft GeV spectra in this class of blazars.  In HBLs, however, it
is the steep high-energy tail of the synchrotron emission that fills
the BAT energy band, yielding {\it soft} BAT spectra.  In these
sources the Compton peak is at TeV energies, beyond the LAT
sensitivity range; thus, at MeV-GeV energies this component is still
rising, and {\it hard} LAT spectra are expected. Figure~2 shows that,
indeed, sources with hard BAT spectra inevitably have soft gamma-ray
spectra, and vice-versa.
  
According to Figures~3 and 4, the position of the Compton peak depends
on the gamma-ray luminosity. The more gamma-ray luminous blazars have
steeper $\Gamma_G$ and flatter $\Gamma_X$, i.e., their Compton peak
tends to fall between MeV and GeV energies. On the contrary, less
luminous GeV sources exhibit the opposite behavior, i.e., they have
Compton peaks beyond the LAT band. We have thus recovered, at the
higher energies and with a {\it high-energy selected blazar sample},
the main result of the blazar sequence -- the luminosity dependence of
the peak energy position. This adds to the recent evidence from the
gamma-rays only (Ghisellini et al. 2009).

Variability will not change significantly the position of FSRQs and
most LBLs in Figure~2. While all sources vary in flux, the most
dramatic continuum shape variations so far have been observed above
the synchrotron peak, i.e., for the X-ray spectra of HBLs. These
objects are expected to move horizontally in the $\Gamma_X$ \gtsima 2
region in the majority of cases, as shown by 1ES0033+595.  Extreme BL
Lacs, which can reach $\Gamma_X \sim 1$, are the only exception but
only a few examples are known so far (Costamante et al. 2001).

Another result of Figure~2 is that a subgroup of objects traditionally
classified as BL~Lacs are located nearby FSRQs but completely set
apart from HBLs. These sources are characterized by similar X-ray
continua as powerful quasar-like blazars, $\Gamma_X \sim 1.8$, but
flatter gamma-ray slopes, $2.1 < \Gamma_G < 2.3$.  Among these sources
we find objects like ON~325 and BL~Lac itself, known for their
``intermediate'' properties. For example, BL~Lac has a concave
0.5--10~keV spectrum (Tanihata et al. 2000) which is intepreted as the
crossing of the synchrotron and Compton components (as predicted by
the sequence for mid luminosities), and which occasionally display
broad emission lines in the optical (Vermeulen et al. 1995). On the
contrary, none of the HBLs was ever observed to exhibit broad optical
lines.
  
Our idea is that the phenomenological definition of BL~Lacs, based on
upper limits to the emission line equivalent width (e.g., Angel \&
Stockman 1980) leads to a ``hybrid'' class, composed of two subgroups:
sources that intrinsically lack broad emission line regions (the {\it
 bona fide} BL~Lacertae objects) and sources that, like BL~Lac
itself, occasionally exhibit broad Balmer lines.  In the lineless
objects the accretion flow is largely sub-Eddington ($\dot{m} \leq
10^{-2}$) and therefore radiates inefficiently so that an optically
thick disk does not form and the broad line region is absent.  They
are TeV sources and are unified with FRI radio galaxies according to
unification schemes.

The objects of the second group are classified differently depending
on the epoch of observation, appearing as quasar-like during low
states of the jet non-thermal continuum, and BL~Lacertae-like when the
jet emission is in a high state and swamps the line emission.  In
these objects the accretion flow is relatively weak, but above the
limit for inefficient accretion ($\dot{m} \geq 10^{-2}$) so that a
broad line region is formed but with sub-Eddington luminosity.  Thus,
it is easier for the non-thermal continuum to wash out the emission
lines in high states.  The presence of such broad line region affects
the jet high-energy spectrum, though less strongly than for ordinary
FSRQs: the radiative losses due to IC scattering of the relativistic
electrons in the jet on the BLR photons can be responsible for a lower
peak energy of the electrons and therefore for a Compton peak at
intermediate energies, thus accounting for an intermediate gamma-ray
spectral index (Celotti \& Ghisellini 2008).  The parent population of
these occasional line emitters should correspond to intermediate
FRI/II morphologies (Kollgaard et al. 1991).
  
From this perspective, Figure~2 shows that the shape of the
high-energy emission offers an independent mean to probe the
astrophysical conditions at the blazar nucleus. The hard X-ray through
gamma-ray continuum carries information on the nuclear ambient where
the jet forms and radiates, being sensitive to the energy density of
photons in and around it, and can be read as a proxy for the accretion
rate (in Eddington units) onto the black hole.  This holds on a
long-term "average". Clearly large flares in the jet involving events
of increased particle acceleration/injection can produce large
variations in the high-energy spectrum. Detailed spectral modeling is
required for a more quantitative assessment of the fundamental
properties (jet power, accretion rate, and mass) of these blazars
(Ghisellini \& Tavecchio 2008).

In conclusion, using a high-energy selected blazar sample we have
recovered one of the main tenets of the so-called blazar sequence --
the luminosity dependence of the position of the Compton peak. The
high-energy emission appears as a promising tool to differentiate
among blazar flavors in an unbiased way. From this perspective, we
look forward to many more years of BAT and LAT synergy.

\acknowledgements  
  
We thank the BAT and LAT teams for making these observations
possible. Interesting conversations with C. Dermer and M. Kadler in
earlier phases of the work are also acknowledged. An anonymous referee
raised some good points in his/her report that stimulated additional
work.

\section{Appendix} 

As can be seen in Figure~7 of Abdo et al. (2009a), the LAT is
inherently biased towards steep-spectrum/bright sources.  One can thus
wonder whether these biases affect the trends between spectral index
and luminosity observed in Figure~3. On the other hand no strong
selection effect has been reported for BAT (e.g., Ajello et al. 2008,
Tueller et al. 2009). 

In order to test this possibility, we performed a detailed Monte-Carlo
study explicitly including the sensitivity limits of the detectors. We
start by recasting Figure~3a-b into a unified plot by plotting the
difference of the BAT and LAT indices,
$\Delta\Gamma=\Gamma_{X}-\Gamma_{G}$, versus the luminosity for
semplicity of analysis; the result is shown in Figure~5. As expected,
more luminous sources have more negative $\Delta\Gamma$. Performing a
linear correlation analysis we find a coefficient of variation, i.e.,
the squared of the linear correlation coefficient, of 0.57 between
$\Delta\Gamma$ and the luminosity in Figure~5.

The main goal of the simulation is to verify whether the trend
observed in Figure~5 is real or due to instrumental biases.  The
blazar sequence manifest itself as a correlation between luminosity
and spectral shape (in our case photon index). The lower the
luminosity, the flatter the photon index (i.e. the high-energy peak
shifts at larger energies) while the larger the luminosity the steeper
the index (e.g., $\Gamma_{G}>$2.3). In this set of Monte Carlo
simulations we aim at generating samples of sources where luminosity
and indices are uncorrelated, corresponding to the case in which the
blazar sequence does not exists, and a-posteriori applying the
selection effects of LAT. If out of a large number of simulations we
found that the correlation index is \gtsima 0.57 for a significant
number of cases, the trend in Figure~5 would be produced, very likely,
by instrumental effects.

For this set of simulations we drew random sources from the flux,
index, and redshift distributions observed for the LAT sources (see
Abdo et al. 2009a).  We remark that the distributions are sampled
indipendently in order to study whether the LAT biases would
a-posteriori introduce any correlation.  Moreover, in order to be as
accurate as possible, the sources were not drawn from a model fit to
the observed distributions (e.g., a Gaussian fit to the index or
redshift distribution), but rather from the raw observed distributions
themselves.  From a physical point view, in this sample low-luminosity
hard FSRQs and very luminous high-redshift BL Lac objects exist in
comparable numbers.  We then rejected all those sources which given a
certain flux would have a photon index too steep to be detected by LAT
(see Fig.7 of Abdo et al. 2009a).  To each LAT source we coupled a BAT
photon index extracted indipendently from the observed distribution of
BAT photon indices, the latter described by a Gaussian centered at
$\Gamma_{BAT}=1.86$ and with dispersion 0.46. Then, we grouped the
sources in samples counting 21 objects and produced the corresponding
distribution shown in Figure~5.  Only 34 out of 1000 plots have a
correlation index larger than 0.57, implying that the probability of
observing by chance Figure~5 correlation is only 3.4\%.

In a sense, LAT cannot observe (if they existed) low luminosity FSRQs,
but can easily detect luminous BL Lac objects. If that were the case
(i.e., there are BL Lacs at high redshifts), no correlation would exist
between the LAT luminosity and difference between the BAT and the LAT
photon indices. While it is not within the scope of this paper to
discuss the existence of BL Lac objects at very high redshift we
conclude noting that these objects are currently not present in the
LAT datasets although not all BL Lac objects have a redshift
measurement available at the moment (see Abdo et al. 2009a).


  
\begin{figure}[h]  
\centerline{\psfig{figure=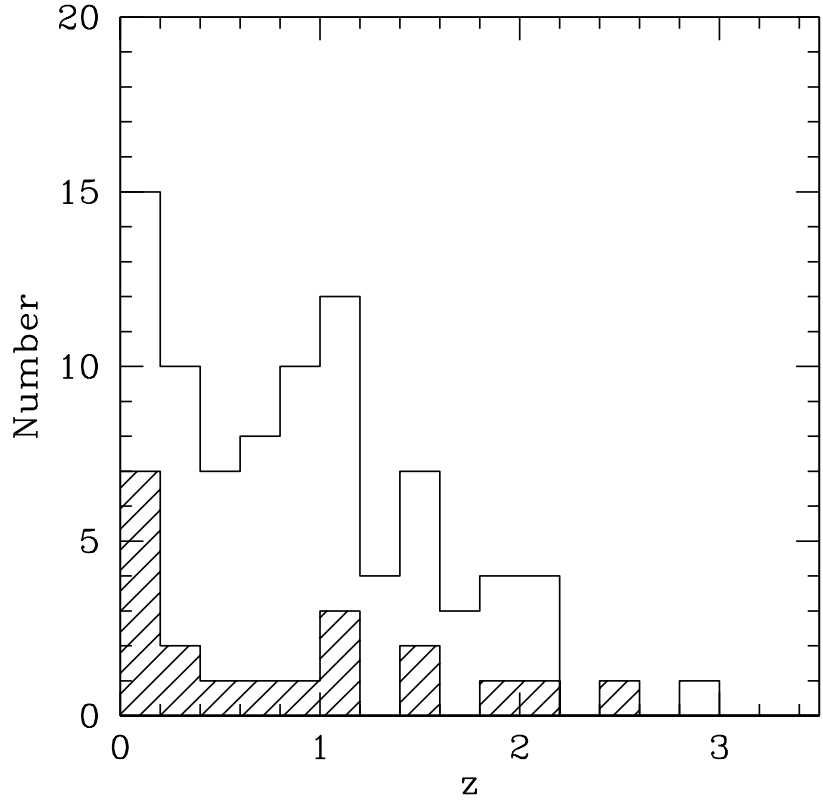,height=5.in}}  
\caption{\footnotesize Distributions of the redshifts of the blazars
  detected with both the BAT and LAT (shaded histogram) and of the LAT
  3 months blazar list (Abdo et al. 2009a). 
 }
\end{figure}  
  
\newpage

  
\begin{figure}[h]  
\centerline{\psfig{figure=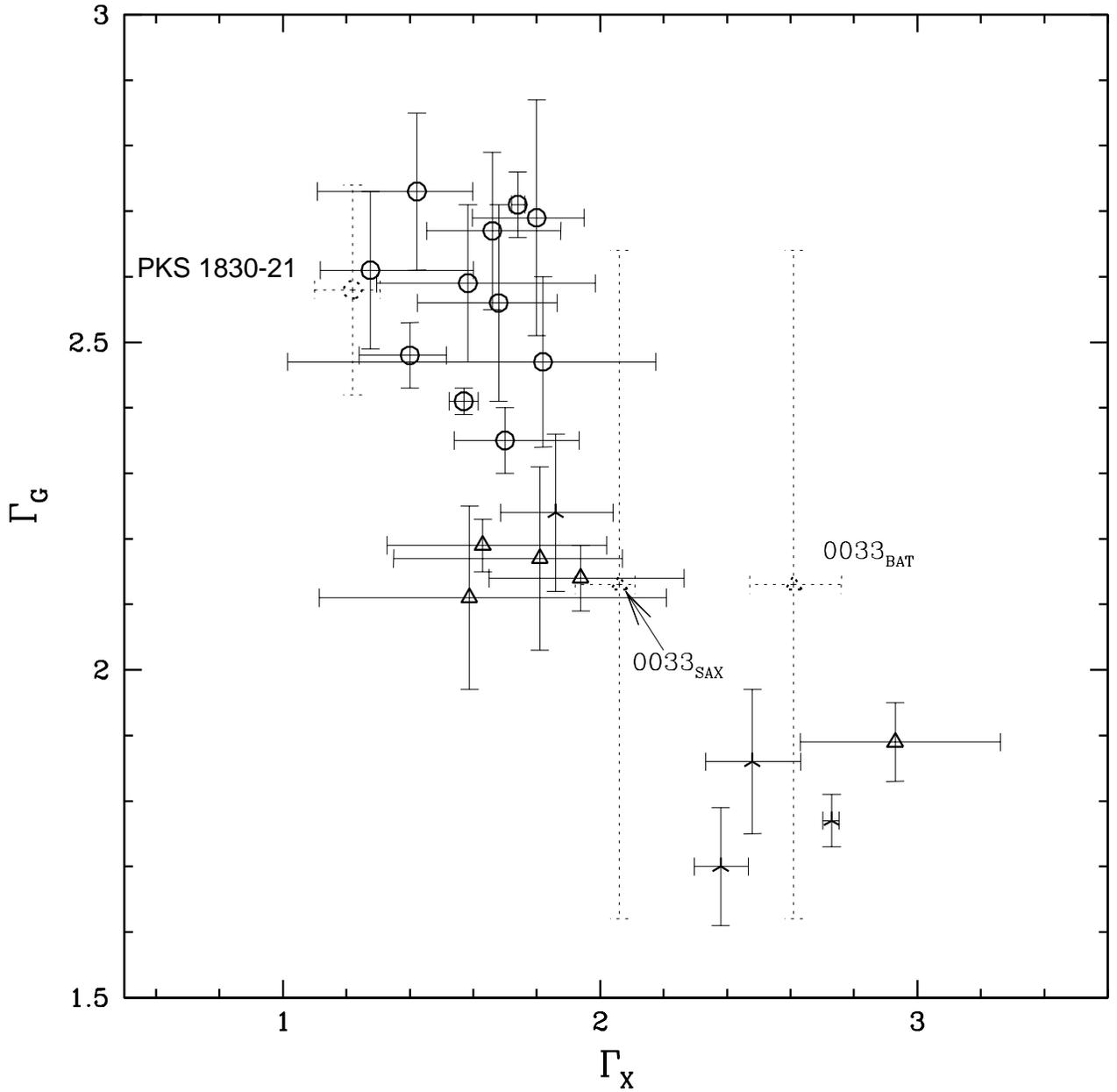,height=7.in}}  
\caption{\footnotesize Plot of the LAT photon index, $\Gamma_G$,
  versus the BAT photon index, $\Gamma_X$, for the blazars detected by
  both instruments (see text). Open circles are FSRQs, open triangles
  are BL Lacs, and starred triangles are BL Lacs detected at TeV
  energies.  Errorbars are 1$\sigma$.  }
\end{figure}  
  
\newpage  
  
  
\begin{figure}[h]  
\centerline{\psfig{figure=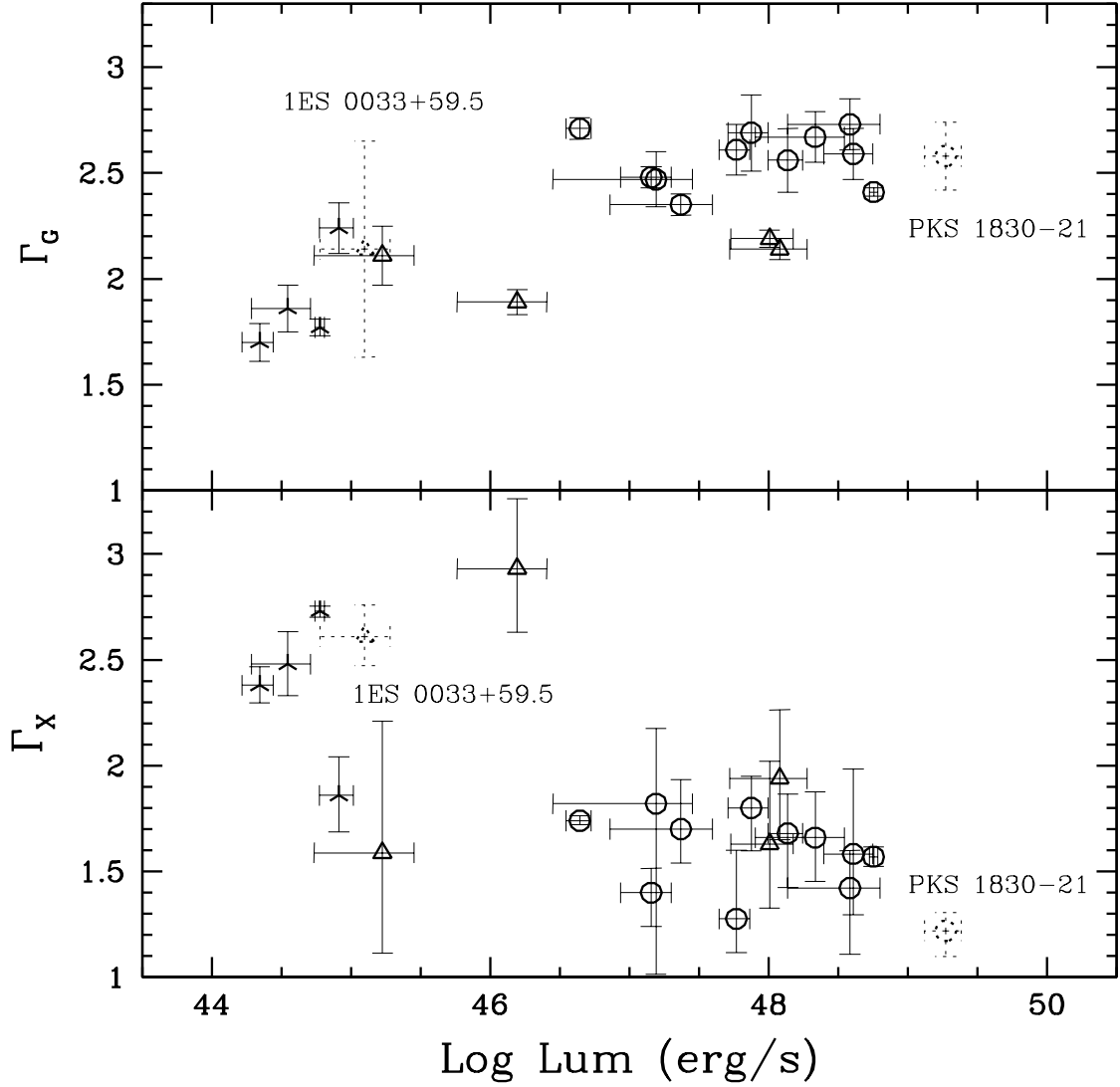,height=7.in}}  
\caption{\footnotesize {\it (a, Top)} Plot of the LAT photon index
  versus the LAT luminosity (from Abdo et al. 2009, Ghisellini et
  al. 2009), showing a trend of steeper slope for more luminous
  sources. {\it (b, Bottom)} Plot of the BAT photon index versus the
  BAT luminosity showing the opposite behavior than in (a), i.e., a
  flatter X-ray continuum for increasing luminosity.  Open circles are
  FSRQs, open triangles are BL Lacs, and starred triangles are BL Lacs
  detected at TeV energies. Errorbars are 1$\sigma$. }
\end{figure}

\newpage  
  
  
\begin{figure}[h]  
\centerline{\psfig{figure=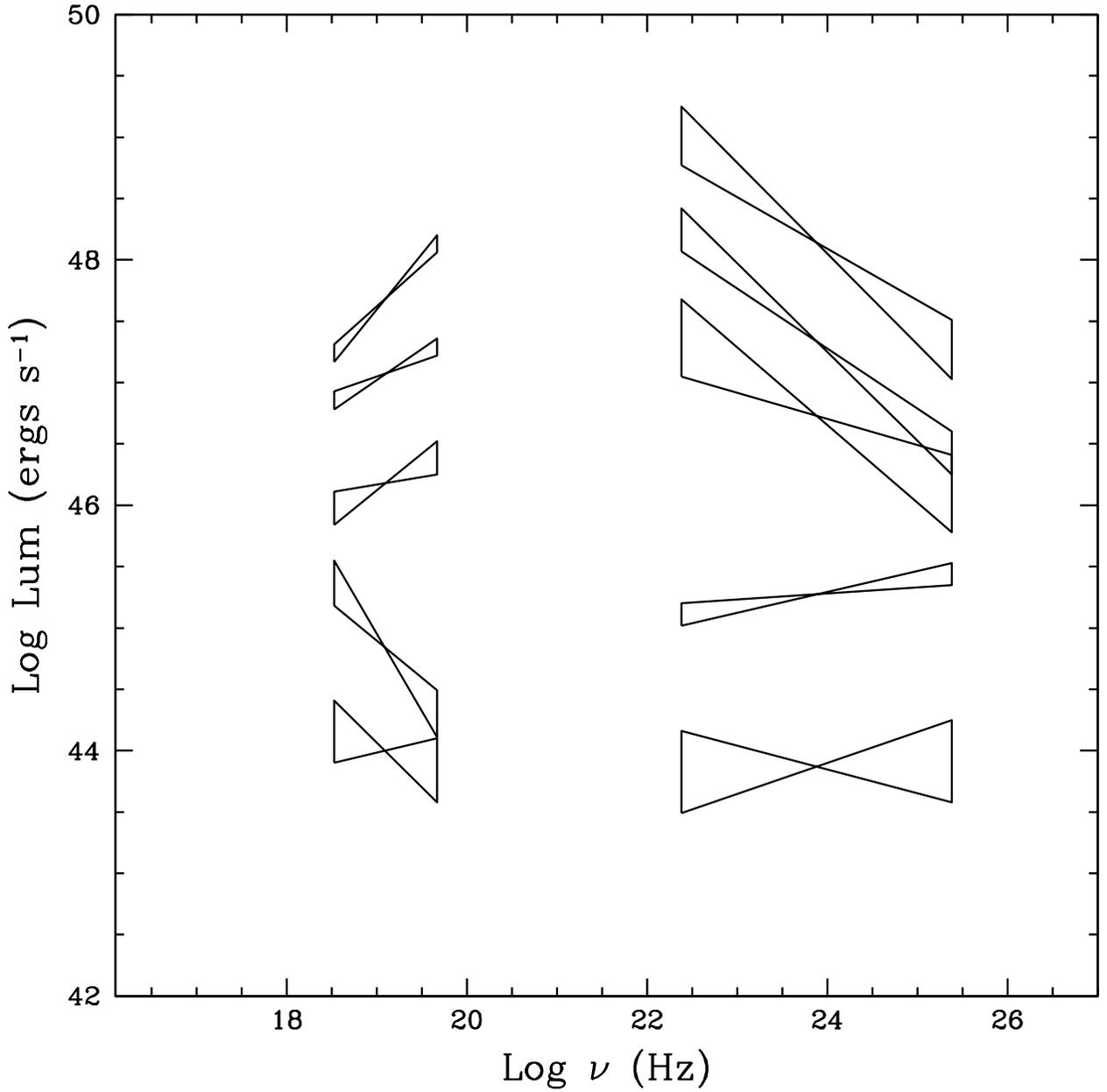,height=7.in}}  
\caption{\footnotesize Plot of the BAT and LAT average continua in
  bins of BAT luminosity (see text). This shows that the IC peak shifts to
  lower energies with increasing luminosity, recovering the Fossati et
  al. (1998) ``blazar sequence''. Errorbars are 1$\sigma$. }
\end{figure}

\newpage  
  
  
\begin{figure}[h]  
\centerline{\psfig{figure=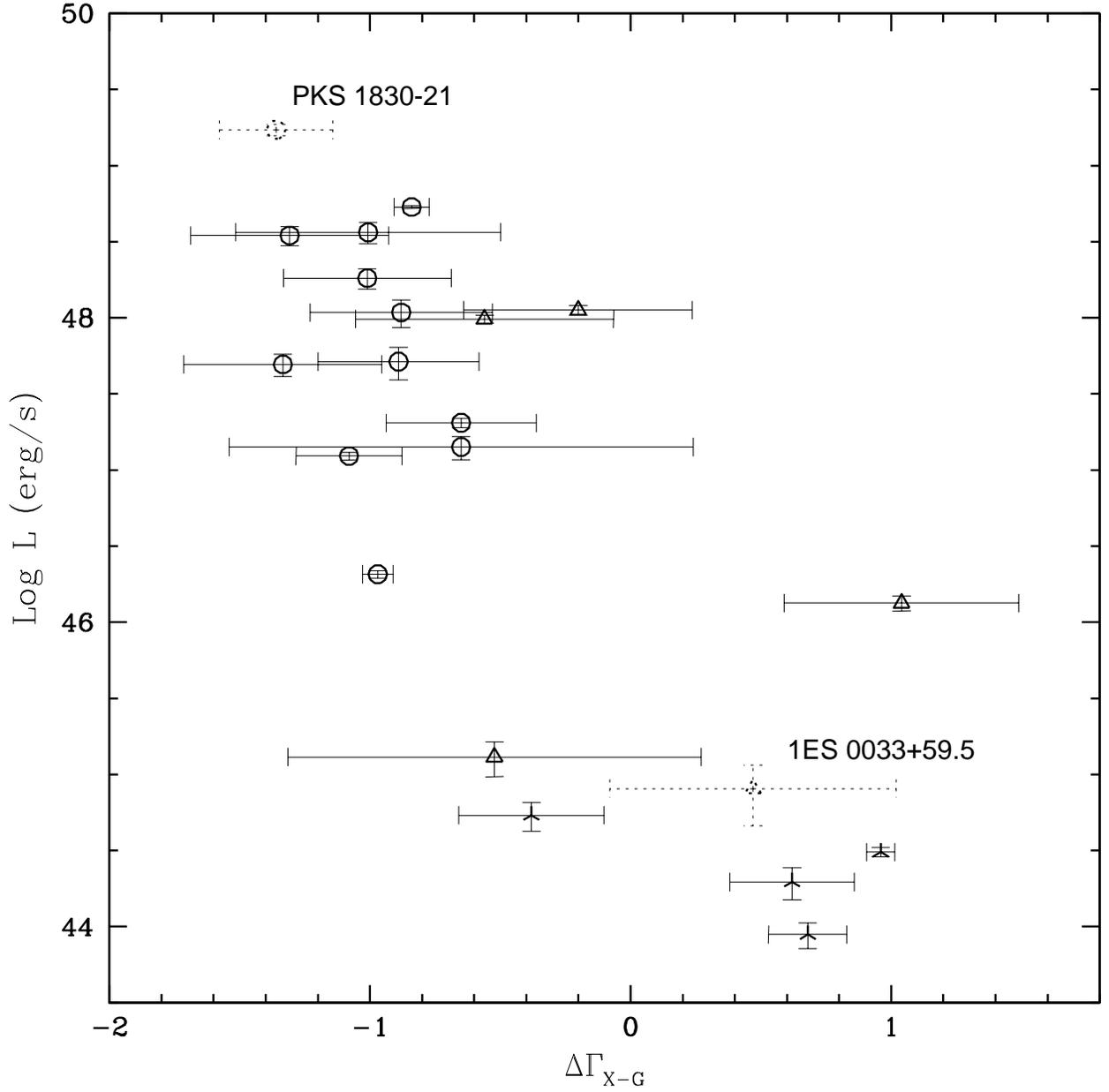,height=7.in}}  
\caption{\footnotesize Plot of the difference between the BAT and LAT  
  slopes versus the LAT luminosity. Open circles are FSRQs, open  
  triangles are BL Lacs, and starred triangles are BL Lacs detected at  
  TeV energies. Errorbars are 1$\sigma$. }  

\end{figure} 
  
\end{document}